\documentclass[letterpaper, 10 pt, journal, twoside]{IEEEtran}
\usepackage{graphics} % for pdf, bitmapped graphics files
\usepackage{epsfig} % for postscript graphics files
\usepackage{amsmath} % assumes amsmath package installed
\usepackage{amssymb}  % assumes amsmath package installed
\usepackage{algorithm}
\usepackage{mathrsfs}
\usepackage{mathtools}
\usepackage{multirow}
\usepackage{comment}
\usepackage{color}
\usepackage{cite}
\usepackage{url}
\usepackage{soul}
\usepackage{listings}
\usepackage[utf8]{inputenc}
\usepackage[T1]{fontenc}
\usepackage[english]{babel}
\usepackage{xcolor}
\usepackage[normalem]{ulem}

\usepackage{xpatch} % also loads expl3
\usepackage{subcaption}

\usepackage{graphicx}
\usepackage{graphics}
\usepackage{amssymb}
\usepackage{amsmath}
\usepackage{color}
\usepackage{xspace}
\usepackage{algpseudocode}
\usepackage{bbm}
\usepackage{comment}
\usepackage{algorithm}
\usepackage{algorithmicx}
\usepackage{soul}
\usepackage{array}
\usepackage[bottom]{footmisc}
\usepackage{tikz}
\usetikzlibrary{shapes,arrows}
\usetikzlibrary{decorations.pathreplacing,decorations.markings,decorations.pathmorphing,decorations.shapes}
\usetikzlibrary{positioning,fit}
\usetikzlibrary{calc}
\usepackage{multirow}
\usepackage{cancel}
\usepackage{hyperref} 
\usepackage{xr}
\tikzstyle{block}=[draw opacity=0.7,line width=1.4cm]

\definecolor{CranJ}{cmyk}{0,0.69,0.54,0.04} %cranberry jello
\definecolor{PinkJ}{cmyk}{0,0.71,0.43,0.12} %pink jeep
\definecolor{Cran}{cmyk}{0,0.73,0.41,0.29} %cranberry 
\definecolor{VRed}{cmyk}{0,0.75,0.25,0.2} %violetred
\definecolor{ORed}{cmyk}{0,0.75,0.75,0} %orangered4
\definecolor{CBlue}{cmyk}{1,0.25,0,0} %curacao	
\setstcolor{VRed}
\usepackage{bm}

\usepackage{caption}

\newlength\myindent


\tikzset{cloud/.pic={
\node[cloud, cloud puffs=10.8,cloud puff arc=110, aspect=2, draw, text width=3cm
    ] () at (0,0) {\tikzpictext};
}}

\newcommand{\real}{{\mathbb{R}}}

\newcommand{\vect}[1]{\boldsymbol{\mathbf{#1}}}

\newcommand{\oprocendsymbol}{\hbox{$\bullet$}}
\newcommand{\oprocend}{\relax\ifmmode\else\unskip\hfill\fi\oprocendsymbol}

\allowdisplaybreaks

\makeatletter
\renewcommand*{\@opargbegintheorem}[3]{\trivlist
      \item[\hskip \labelsep{\emph{ #1\ #2}}] \emph{(#3):}\ \itshape}
\makeatother

\usepackage{tcolorbox}% http://ctan.org/pkg/tcolorbox
\definecolor{mycolor}{rgb}{0.122, 0.435, 0.698}% Rule colour
\makeatletter
\newcommand{\mybox}[1]{%
  \setbox0=\hbox{#1}%
  \setlength{\@tempdima}{\dimexpr\wd0+13pt}%
  \begin{tcolorbox}[colframe=mycolor,boxrule=0.5pt,arc=4pt,
      left=6pt,right=6pt,top=6pt,bottom=6pt,boxsep=0pt,width=\@tempdima]
    #1
  \end{tcolorbox}
}
\makeatother
\usepackage{cite}
\makeatletter
\def\hlinewd#1{%
\noalign{\ifnum0=`}\fi\hrule \@height #1 \futurelet
\reserved@a\@xhline}
\makeatother

\begin{document}

\title{NavEX: A Multi-Agent Coverage in Non-Convex and Uneven Environments via Exemplar-Clustering 
%Maximal Spatial Stein Variational Coverage for multisensor  Deployment
}

\author{Donipolo Ghimire${^1}$~~ Carlos Nieto-Granda${^2}$ ~~ Solmaz S. Kia${^1}$, \emph{Senior Member, IEEE}
\thanks{${^1}$D.~Ghimire and S.~Kia are with the Department of Mechanical and Aerospace Engineering, University of California, Irvine, CA 92697 {\tt\footnotesize dghimire,solmaz@uci.edu}} 
\thanks{${^2}$ C.~Nieto is with the U.S. DEVCOM Army Research Laboratory (ARL), Adelphi, MD 20783 {\tt\footnotesize,carlos.p.nieto2.civ@army.mil}}}
% \thanks{Digital Object Identifier (DOI): see top of this page.}}

% \markboth{IEEE Robotics and Automation Letters. Preprint Version. March, 2024}
% {Ghimire \MakeLowercase{\textit{et al.}}: Dispatch Coverage} 

% \markboth{IEEE Robotics and Automation Letters. Preprint Version. }
% { \MakeLowercase{\textit{et al.}}: Monitoring using UAV}

\maketitle

%%%%%%%%%%%%%%%%%%%%%%%%%%%%%%%%%%%%%%%%%%%%%%%%%%%%%%%%%%%%%%%%%%%%%%%%%%%%%%%%
%====================
% Abstract
%====================
\begin{abstract}
This paper addresses multi-agent deployment in non-convex and uneven environments. To overcome the limitations of traditional approaches, we introduce Navigable Exemplar-Based Dispatch Coverage (NavEX), a novel dispatch coverage framework that combines exemplar-clustering with obstacle-aware and traversability-aware shortest distances, offering a deployment framework based on submodular optimization. NavEX provides a unified approach to solve two critical coverage tasks: (a) fair-access deployment, aiming to provide equitable service by minimizing agent-target distances, and (b) hotspot deployment, prioritizing high-density target regions. A key feature of NavEX is the use of exemplar-clustering for the coverage utility measure, which provides the flexibility to employ non-Euclidean distance metrics that do not necessarily conform to the triangle inequality. This allows NavEX to incorporate visibility graphs for shortest-path computation in environments with planar obstacles, and traversability-aware RRT$^\star$ for complex, rugged terrains. By leveraging submodular optimization, the NavEX framework enables efficient, near-optimal solutions with provable performance guarantees for multi-agent deployment in realistic and complex settings, as demonstrated by our simulations.
\end{abstract}

\begin{IEEEkeywords}
 Multi-agent deployment; coverage; non-convex; uneven; submodular maximization.
\end{IEEEkeywords}
%====================
% Introduction
%====================

\section{Introduction}
\vspace{-0.05in}
\IEEEPARstart{A}{}rea coverage is a fundamental problem in applications such as environmental monitoring, infrastructure surveillance, and resource allocation in robotic networks~\cite{SSK-SM:24,ME-RG-DD:11}. Effective coverage necessitates strategically positioning assets—such as sensors, stations, or robots—within a spatial two- or three-dimensional workspace, denoted by $\mathcal{W}$, 
  to maximize a specific coverage utility measure. The coverage utility is typically represented by a function that quantifies detection capability, information acquisition, or facility location accessibility for a timely spatial response. These utility functions are often defined based on the distance between target points and the deployment location, weighted by a distribution that reflects the spatial spread of targets or an area priority function. To formalize our discussion, let us define the weighted distance as $d(q, p_{i})$, where $q\in\mathcal{W}$ is a point in an area and $p_i\in\mathcal{X}$ is the location of asset $i\in\mathcal{A}=\{1,\cdots,N\}$, $N$ being the total number of assets. Here, $\mathcal{X}\subseteq\mathcal{W}$ is the set of admissible deployment locations. 

Most existing coverage algorithms assume a convex environment and base the utility function on Euclidean distances $d(q, p_{i})=\|q-p_i\|$. In these approaches, the area is typically divided into sub-regions, with agents assigned to each, thereby ensuring effective spatial coverage~\cite{JC-SM-FB:04}. However, when the environment is non-convex and uneven, the concept of distance must respect constraints like obstacles, traversability conditions and irregular boundaries, complicating the definition of the utility function and even the mathematical description of the target distribution or area priority function. In a continuous space, non-convexity introduces distortion of the distance metric, making it difficult to model the cost or feasibility of covering the targets. In this paper, we aim to solve coverage problems by designing an appropriate utility function for scenarios within a non-convex environment, where $\mathcal{X}$ is a discrete set of pre-specified deployment locations.

\emph{Related literature}: 
Voronoi partitioning is widely used in mobile task allocation, exploration, and sensing networks. Initially developed for convex domains~\cite{OA-19,JC-SM-FB:04,MS-DR-JS:09}, it has been adapted for non-convex domains. For instance, one study~\cite{CN-CH-ZM:08} approximated non-convex domains using diffeomorphism, while others~\cite{YS-MT-AT:15,SB-RG_VK:14} used geodesic distance instead of Euclidean distance, allowing regions to extend into non-visible areas. Additionally, Voronoi partitioning has been applied to multi-robot exploration in non-convex domains via local Voronoi decomposition and deep reinforcement learning~\cite{JH-FA:20}. Similarly,~\cite{YA-AA-SS:23} employs a distributed approximation algorithm for coverage in non-convex environments, improving over Lloyd's algorithm, which determines Voronoi cells in discrete environments.
Despite research in defining robust coverage functionals, existing methods for non-convex domains have limitations. High computational complexity of geodesic methods limits scalability, and performance declines in domains with disjoint target areas~\cite{LK-LK:24}.
To address this, discretized coverage approaches represent environments as graphs and solve a k-median problem to minimize agents' service time~\cite{LS-OS:16}. ~\cite{RR-LZ-GS:20,SX-GC:19} used greedy algorithms on discretized graph models to exploit submodularity, achieving approximate guarantees. However, solution quality depends on the graph's ability to represent the environment and approximate distances. By integrating these approaches, our work utilizes appropriate distance metrics on a discretized graph model to achieve effective coverage and computational tractability.

When the workspace $\mathcal{W}$ is convex, and the set of candidate deployment points $\mathcal{X}$ is discrete and pre-specified, and utility functions such as information-theoretic or geometric measures are considered, the deployment problem can be cast as submodular maximization subject to cardinality or partition matroid constraints~\cite{AK-AS-CG:08,SSK-SM:24,kia2025submodular} (see Section~\ref{sec::2025_RAL_Deployment_Method} for submodular function definition). The submodular property enables near-optimal, polynomial-time solutions for these NP-hard problems~\cite{JL-RKW:19,ZX-VT:22,NR-SSK:23auto,GS-GSS:24}. However, in non-convex environments, these methods cannot perform effectively due to the complexity associated with defining the notion of distance required by utility functions, such as the presence of obstacles and non-euclidean distance metrics.

\textit{Statement of Contribution:}  In this letter, we introduce NavEX (Navigable Exemplar-Based Dispatch Coverage), a flexible framework for optimizing facility location in constrained, nonconvex, and uneven environments. Our approach integrates: \emph{Exemplar-clustering}~\cite{KL-PR:90,BA-MB-KA:14} and shortest distance measures using visibility graphs or traversability-aware RRT$^\star$~\cite{MD:00,JG-SS-TB:14}. By combining these approaches, NavEX framework accommodates constraints such as obstacles and traversability. By showing that this framework is an instance of submodular maximization under a matroid condition, we present a polynomial-time solution with characterizable optimality gap. We apply the NavEX framework for two problems that want to position agents strategically while considering obstacles and traversability:
\begin{itemize}
    \item \textit{Fair-Access Deployment Problem}:  The goal is to minimize the maximum distance between any target point in $\mathcal{W}$ and its nearest agent, ensuring equitable access.
    
\item \textit{Hotspot Deployment Problem}: The goal is to deploy in the most densely populated or high-demand areas in $\mathcal{W}$. 
\end{itemize}

The fair-access deployment problem is relevant in scenarios such as emergency response and public service distribution, where equitable access is critical. The hotspot deployment problem addresses scenarios focusing on regions with the highest concentration of potential users or events. Both problems necessitate non-Euclidean distance metrics to account for environmental constraints and require deployment strategies that consider population distribution and traversability. The proposed NavEX framework bridges theoretical models and practical applications, offering robust solutions for facility location in real-world conditions. Specifically, our contribution is a unified framework that addresses both fair-access and demand-based deployment, advancing coverage optimization in challenging spaces. Simulation results demonstrate the effectiveness of our approach. The code used in these simulations is available in our GitHub repository\footnote{Please visit \href{https://github.com/donipologhimire/MultiAgentCoverageviaExemplarClustering}{github.com/donipologhimire/NavEx}.}

\section{Problem setting}
\label{sec::prob}
\vspace{-0.05in}
We consider a multi-agent deployment problem where the goal is to solve either the fair-access or hotspot deployment tasks. These problems require positioning $K\in\mathbb{Z}_{>1}$ agents among $M\in\mathbb{Z}_{>1}$ targets scattered over a finite space $\mathcal{W}\subset\real^3$, where $\mathcal{W}$ is either a flat surface with solid obstacles or an uneven space, which limits traversability conditions for the deployed agents. The candidate discrete deployment points $\mathcal{X}\subset \mathcal{W}$, $|\mathcal{X}|>K$, are prespecified. The set of target locations is give by $\mathcal{T}=\{\vect{x}^T_1,\cdots,\vect{x}^T_M\}$, where $\vect{x}^T_i\in\mathcal{W}$ is location of target $i\in\{1,\cdots,M\}$. The targets can be actual entities such as people or any other service requester or virtual points sampled based off of area priority or information distribution maps. Moreover, topological map of $\mathcal{W}$ is given from prior~surveys. 

The presence of obstacles and traversability limitations renders traditional Euclidean or geodesic distance metrics inadequate for accurately characterizing the spatial relationships between agents and targets. Therefore, the objective of this paper is to develop a comprehensive framework comprising appropriate distance metrics and an optimization methodology tailored to these metrics, enabling the effective solution of both fair-access and hotspot deployment tasks.

\section{NavEX: a multi-agent deployment in non-convex and uneven workspaces}
\label{sec::2025_RAL_Deployment_Method}
\vspace{-0.05in}
In NavEX, we cast the deployment problem given a set of prespecified deployment candidate points, as a set-function maximization problem. When there is a single authority in charge of deploying $K$ agents, the problem is formulated as  
\begin{equation}\label{eq::deploy_uniform}
\max_{\mathcal{S}\subset\mathcal{X}} f(\mathcal{S})~~\text{s.t.}~~|\mathcal{S}|\leq K.
\end{equation}
where $f:2^{\mathcal{X}}\to\real_{\geq0}$ is the utility measure quantifying the coverage objective. 

Similarly in NavEX, we can also consider a multi-authority variation of the deployment problem, where $n\in\mathbb{Z}_{>1}$ authorities each want to deploy $\kappa_i\in\mathbb{Z}_{\geq} 1$ agents at subsets $\mathcal{X}_i\subset\mathcal{X}$ where $|\mathcal{X}_i|>\kappa_i$ and $\sum_{i=1}^n\kappa_i=K$ and $\cup_{i=1}^n\mathcal{X}_i=\mathcal{X}$. This multi-authority deployment problem can be formulated as:
\begin{equation}\label{eq::deploy_partition}
\max_{\mathcal{S}\subset\mathcal{X}} f(\mathcal{S})~~\text{s.t.}~~|\mathcal{S}\cap\mathcal{X}_i|\leq \kappa_i,~~ i\in\{1,\cdots,n\}.
\end{equation}
This formulation allows for the consideration of multiple authorities with distinct deployment constraints while still optimizing the overall utility function. 

The optimization problems of type~\eqref{eq::deploy_uniform} or \eqref{eq::deploy_partition} are often NP-hard unless for very few utility functions. However, when the utility function is monotone increasing and submodular\footnote{A set function $f:2^{\mathcal{X}}\to\mathbb{R}_{\geq0}$ is monotone increasing if for any $\mathcal{S}\subset\mathcal{R}\subset\mathcal{X}$ we have $f(\mathcal{S})\leq f(\mathcal{R})$. Moreover, the function is submodular if and only if for any $\mathcal{S}\subset\mathcal{R}\subset\mathcal{X}$ and $p\in\mathcal{X}$ we have $f(\mathcal{S}\cup\{p\})-f(\mathcal{S})\geq f(\mathcal{R}\cup\{p\})-f(\mathcal{R})$. Marginal gain $f(\mathcal{S}\cup\{p\})-f(\mathcal{S})$ often is called the \emph{discrete derivative} of $f$ at $\mathcal{S}$ with respect to $p$. Marginal gain measures the additional value or utility obtained by including the element $p$ into the set $\mathcal{S}$. Marginal gain helps in understanding the benefit of including an additional element in a set given the current context, i.e., the elements already selected in the set $\mathcal{S}$. These submodular functions exhibit property of diminishing return, see~\cite{kia2025submodular}.}, the so-called \emph{sequential greedy algorithm} presented in Algorithm~\ref{alg::SQG_uniform} and~\ref{alg::SQG_partition} can effectively solve~\eqref{eq::deploy_uniform} and~\eqref{eq::deploy_partition} in polynomial time with respective optimality gaps of $(1-\frac{1}{\text{e}})$ and $\frac{1}{2}$, see e.g.,~\cite{kia2025submodular}. The challenge to utilize this optimization framework for the fair-access or hotspot deployment lies in defining a utility function that is both submodular and compatible with distance metrics suitable for constrained, nonconvex, and uneven environments outlined in Section~\ref{subsec:2025_RAL_distance_metrics}.

\begin{algorithm}[t]
\vspace{0.01in}
 { \small  
\caption{Sequential greedy algorithm for single-authority deployment}

\textbf{Input:} Ground set $\mathcal{X}$ and utility function $f$, number of choices $k$ \\
\textbf{Output:} Subset $\mathcal{S} \subseteq \mathcal{X}$ satisfying $|\mathcal{S}| \leq K$
\label{alg::SQG_uniform}
\begin{algorithmic}

\State $\bar{\mathcal{S}}_0 \gets \emptyset$
\For{$i \in \{1, \dots, K\}$}
    \State $s^* \gets \arg\max_{s \in \mathcal{X} \setminus \bar{\mathcal{S}}_{i-1}} (f(\bar{\mathcal{S}}_{i-1}\cup\{s\})-f(\bar{\mathcal{S}}_{i-1}))$
    \State $\bar{\mathcal{S}}_i \gets \bar{\mathcal{S}}_{i-1} \cup \{s^*\}$
\EndFor
\State $\mathcal{S} \gets \bar{\mathcal{S}}_K$
\end{algorithmic} 
}
\end{algorithm}

\begin{algorithm}[t]
 {\small   
   \caption{Sequential greedy algorithm for multi-authority deployment}

\textbf{Input:} Ground set $\mathcal{X}_1,\cdots,\mathcal{X}_n$ and utility function $f$, number of choices $\kappa_1,\cdots,\kappa_n$ \\
\textbf{Output:} Subset $\mathcal{S} \subseteq \mathcal{X}$ satisfying $|S\cap \mathcal{X}_i| \leq \kappa_i$ , $i\in\{1,\cdots,n\}$
\label{alg::SQG_partition}
\begin{algorithmic}

\State $\bar{\mathcal{S}}_0 \gets \emptyset$
\For{$i \in \{1, \dots, n\}$}
\State $\hat{\mathcal{S}}_i\gets \bar{\mathcal{S}}_{i-1}$
\For{$j\in\{1,\cdots,\kappa_i\}$}
    \State $s^* \gets \arg\max_{s \in \mathcal{X}_i \setminus \hat{\mathcal{S}}_{i}} (f(\hat{\mathcal{S}}_{i}\cup\{s\})-f(\hat{\mathcal{S}}_{i}))$
    \State $\hat{\mathcal{S}}_i \gets \hat{\mathcal{S}}_{i} \cup \{s^*\}$
    \EndFor
    \State $\bar{\mathcal{S}}_i\gets \hat{\mathcal{S}}_{i}$
\EndFor
\State $\mathcal{S} \gets \bar{\mathcal{S}}_n$

\end{algorithmic} 
}
\end{algorithm}

\begin{figure}[t]
    \centering
    \includegraphics[width=0.6\linewidth]{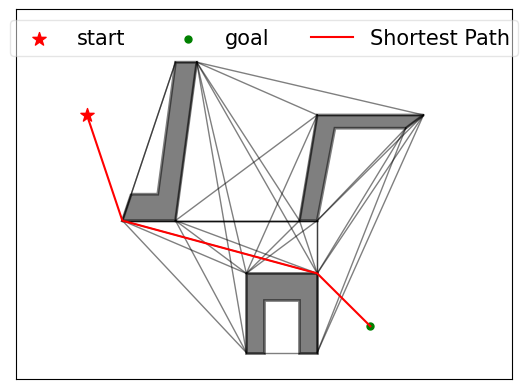}
    \vspace{-0.05in}\caption{{\small A non-convex environment overlaid with the corresponding visibility graph that is used as a distance metric. The shortest path is highlighted in red between two random points.}}
    \label{fig:2025_RAL_visibility_graph_path}
\end{figure}

\begin{figure*}[t]
    \centering
    \begin{subfigure}[t]{0.40\linewidth}
        \centering
        \includegraphics[scale=0.1]{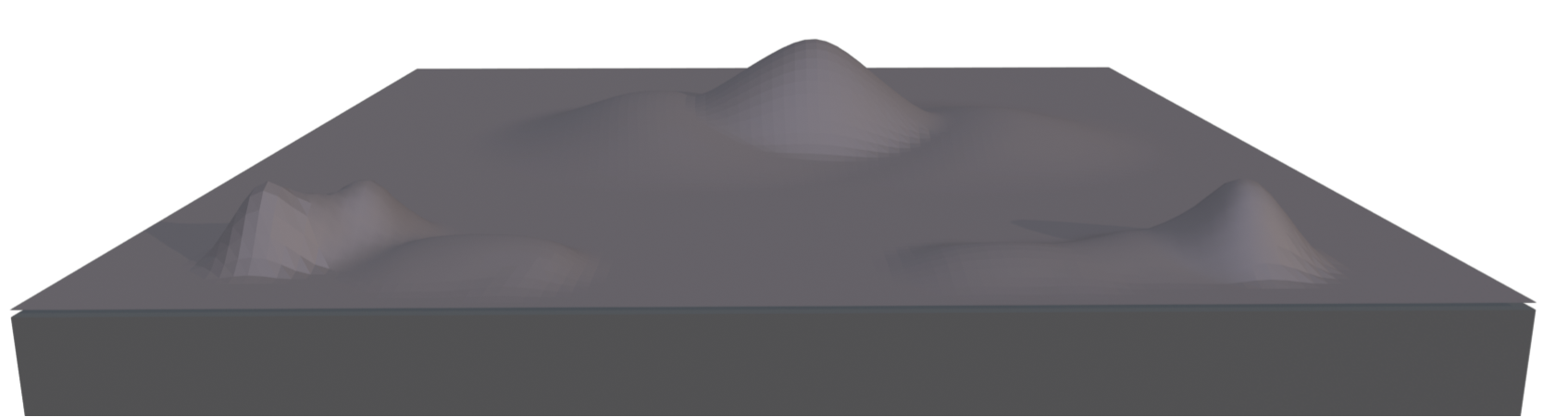}
        \caption{{\small An example of terrain where agents are deployed.}}
        \label{subfig:2025_RAL_fig1a}
    \end{subfigure}
    \hfill
    \begin{subfigure}[t]{0.25\linewidth}
        \centering
        \includegraphics[width=0.9\textwidth]{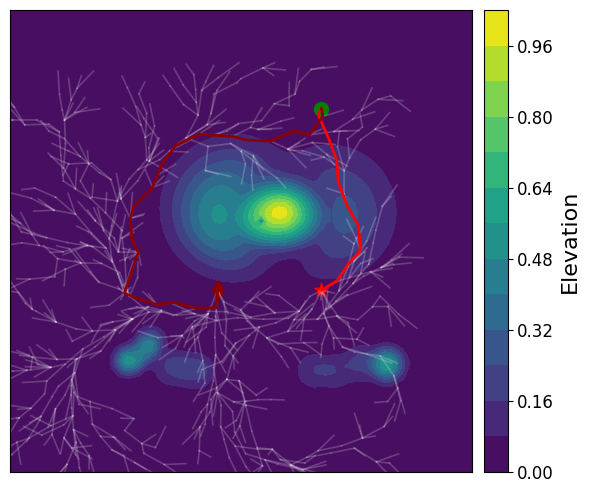}
        \caption{{\small Elevation map with RRT$^{\star}$.}}
        \label{subfig:2025_RAL_elevation}
    \end{subfigure}
    \hfill
     % \vspace{0.5em}  % Adjust spacing between top and bottom rows
    \begin{subfigure}[t]{0.25\linewidth}
        \centering
        \includegraphics[width=0.9\textwidth]{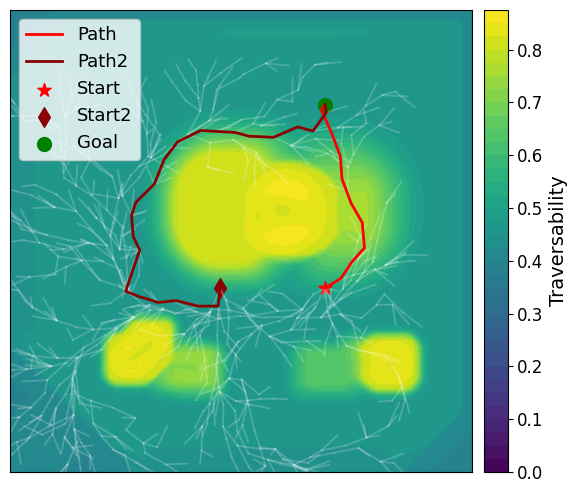}
        \caption{Traversability map with RRT$^{\star}$.}
        \label{fig:2025_RAL_traversability_path}
    \end{subfigure}

    \caption{{\small Distance metric via traversability-aware RRT$^{\star}$: two generated paths show the traversability impacts the distance between start and goal points. Depending on the terrain, the path may entirely circumvent the obstacle formed by hills, while in other traversable regions, the algorithm allows direct passage.}}
    \label{fig:2025_RAL_traversability_example}
\end{figure*}

\subsection{Distance metrics}
\label{subsec:2025_RAL_distance_metrics}
\vspace{-0.05in}
Distance metrics are crucial for determining the spatial relationships between agents and targets. In the NavEX framework, in 2D or planar environments with polygonal-fitted obstacles, we employ the exact shortest paths computed via visibility graphs~\cite{LP-MW:79}. This method quantifies the minimal distances required for traversal, allowing us to capture the navigational cost imposed by obstacles and irregular boundaries, as illustrated in Fig.~\ref{fig:2025_RAL_visibility_graph_path}. Specifically, visibility graphs construct a collision-free roadmap by connecting the visible convex vertices of the obstacles. A vertex is visible from another vertex if a straight line connecting them lies entirely in the free workspace without intersecting any obstacle edges; the edges of the obstacles are considered part of the roadmap. A unique characteristic of the visibility graph is its roadmap efficiency of $1$, meaning that the shortest paths generated by this roadmap are indeed the true shortest paths in the 2D space in the presence of obstacles. Although initially proposed for path planning, we adapt this method to our deployment framework by linking deployment candidate points and target points to visible vertices on the visibility-based graph. This allows us to use Dijkstra's algorithm or its variants to determine the exact shortest path between any two points. While the visibility graph is optimal, it is particularly suitable for environments with a manageable number of vertices and static obstacle configurations. It is important to note the computational complexity: constructing the visibility graph requires $O(n^2 \log{n})$ time, while Dijkstra's algorithm has a complexity of $O(E+V \log{V})$, where $E$ is the number of edges and $V$ is the number of vertices. When the computational cost of visibility graph-based distance measurement is significant, we can resort to approximate methods such as those using the RRT$^{\star}$ algorithm~\cite{SG:07,SK-EF:11}.
%significant computational challenges for larger graphs and higher-dimensional representations of the environment. \margin{say }

For uneven environments, in NavEX framework, we adopt the approach of Leininger et al.~\cite{AL-ML-LL:24} to construct a traversability map $\mathcal{M}_{\tau}$ that characterizes navigable terrain, as depicted in Fig.~\ref{fig:2025_RAL_traversability_path}. Subsequently, we compute shortest distances using traversability-aware RRT$^{\star}$. This algorithm incrementally generates uniform samples and during tree construction and rewiring, nodes and edges are included only if they satisfy a predefined traversability threshold. This path-planning tree algorithm ensures paths remain within navigable regions when deploying agents in uneven terrain, as shown in Fig.~\ref{fig:2025_RAL_traversability_path}. Navigability is determined by assessing terrain characteristics such as slope, flatness, and elevation change (Fig.~\ref{subfig:2025_RAL_elevation}) relative to the vehicle’s safety limits for roll $\phi$ and pitch $\psi$ angles. The slope $\mathcal{M}_\Delta$ is derived from the gradient of the topological map. Flatness $\mathcal{M}_f$ and step height $\mathcal{M}_\zeta$ are similarly computed. For brevity, we refer readers to~\cite{AL-ML-LL:24} for detailed computation of these values. The slope, flatness, and step height local maps are combined to form the traversability map $\mathcal{M}_{\tau}$ of $\mathcal{W}$, as shown in Fig.~\ref{fig:2025_RAL_traversability_example}, where traversability is represented by a value $\tau\in [0,1]$ and calculated as:
\begin{equation}
    \mathcal{M}_{\tau} = \omega_{1} \dfrac{\mathcal{M}_{\Delta}}{s_{\textit{crit}}} + \omega_2 \dfrac{\mathcal{M}_{f}}{f_{\textit{crit}}} + \omega_3 \dfrac{\mathcal{M}_{\zeta}}{\zeta_{\textit{crit}}}
\end{equation}
where $\omega_1,\omega_2,\omega_3\in\mathbb{R}_{>0}$, $\omega_1+\omega_2+\omega_3=1$. The parameters $s_{\text{crit}}$, $f_{\text{crit}}$, and $\zeta_{\text{crit}}$ represent vehicle-specific critical thresholds for maximum slope, flatness, and step height, respectively. These thresholds can be obtained from manufacturer specifications or experimental testing. The weighting parameters $\omega$'s can be adjusted based on vehicle type and terrain characteristics; for example, a tracked vehicle might have a higher $\omega_1$ to reflect its superior slope performance compared to a wheeled~robot. The traversability-aware RRT$^\star$ is both probabilistically complete and asymptotically optimal~\cite{JG-SS-TB:14, SK-EF:11}. From a computational perspective, RRT$^\star$ construction complexity is $O(n \log{n})$ and path length queries between start/goal points $O(n)$~\cite{SK-EF:11}. This significantly decreases the computational cost in higher dimensions compared to visibility graphs. However, our approach demands computing the pairwise distance between target points and candidate deployment points. This requires generating a tree for each pair of start-goal pair, leading to a computational overhead. To address this, we can leverage the independence of these computations--each tree is independent of the others--and parallelize the process. %While we do not explicitly implement this parallelization, it remains a viable direction for enabling simultaneous evaluation of multiple paths and optimizing computation performance. 

\vspace{-0.06in}
\subsection{Utility function design}
\vspace{-0.06in}
Next, we explain how we integrate the \emph{Exemplar-clustering} concept~\cite{KL-PR:90,BA-MB-KA:14} from the data summarization literature with the shortest distance measures derived from graph-based methods--specifically, visibility graphs for 2D or traversability-aware RRT$^\star$ for 3D environments~\cite{MD:00,JG-SS-TB:14}--to define the utility function $f(\mathcal{S})$ in the NavEX framework. This utility function is used for addressing both the fair-access and Hotspot deployment problems. We will solve via~\eqref{eq::deploy_uniform} in the single authority case and  via~\eqref{eq::deploy_partition} in the multi-authority scenario.

Exemplar-based clustering is a technique that identifies a representative subset of elements, termed exemplars, from a larger dataset. This approach is particularly useful for our deployment problems as it allows us to select optimal agent positions from the set of candidate deployment points. The method aims to minimize the cumulative pairwise dissimilarities between chosen candidate deployments (exemplar) from $\mathcal{X}$ and target points (dataset elements) $\mathcal{T}$. This optimization problem can be formulated as:
\begin{equation}
\label{eq::util_min}
L(\mathcal{S})=\frac{1}{|\mathcal{S}|}\sum\nolimits_{p\in\mathcal{X}}\min_{d\in\mathcal{T}} \text{dist}(p,d),
\end{equation}
for any subset $\mathcal{S}\subset\mathcal{P}$, where $\text{dist}(p,d)\geq 0$, represents the dissimilarity or distance between elements. In exemplar-clustering, we seek a subset $\mathcal{R}\subset\mathcal{P}$ subjected to constraints like cardinality, that minimizes $L$. We recast this minimization as a submodular maximization problem by defining the utility~function 
\begin{align}
\label{eq::util_num}
    f(\mathcal{R}) = L(\{d_0\}) - L(\mathcal{R} \cup \{d_0\}),
\end{align}
where $d_0$ is a hypothetical auxiliary element added to the exemplar space. This utility function~\eqref{eq::util_num} quantifies the reduction in the loss when selecting an active set $\mathcal{R}$, compared to the loss associated with only the phantom placement location. Maximizing this function corresponds to minimizing the loss~\eqref{eq::util_min}. Furthermore, the utility function~\eqref{eq::util_num} is demonstrated to be both submodular and monotonically increasing~\cite{RG-AK:10}.

Of significant importance for our deployment problem is the flexibility of the distance measure $\text{dist}(.,.)$ in the exemplar-clustering function~\eqref{eq::util_min}. For the utility function $f(\mathcal{R})$ in~\eqref{eq::util_num} to be monotone increasing and submodular, the distance measure need not be symmetric nor adhere to the triangle inequality. In practice, approaches such as visibility graph or RRT$^\star$, which incorporate traversability and non-euclidean distances yield positive distance measures but do not necessarily conform to triangle inequality. Despite this, as discussed previously, the NavEX framework provides the flexibility to frame our problem as submodular maximization problem subjected to matroid constraint. Consequently, we can solve these problems via Algorithm~\ref{alg::SQG_uniform} or Algorithm~\ref{alg::SQG_partition}. Therefore in NavEX framework, we effectively address the complexities of deployment in non-convex and uneven spaces while maintaining the mathematical properties necessary for efficient optimization. 

For the fair-access deployment problem, we use the shortest distance obtained from either a visibility graph or traversability-aware RRT$^\star$. On the other hand for the hotspot deployment, we will employ a truncated logarithmic transformation of the distance:
\begin{equation}
\label{eq::hotspot_deploy}
    \text{dist}_{\text{Hotspot}}(.,.)=
\begin{cases}
\log({1+\text{dist}(x,y)}) & \text{if } \text{dist}(x,y) \leq \ell, \\[1mm]
L & \text{if } \text{dist}(x,y) > \ell,
\end{cases}
\end{equation}
where, $\ell$ is a threshold and $L$ is a large constant. This transformation is positive--close points yield a lower value and farther points yield a higher value, up to the cap $L$. Moreover, $log(d+1)$ is concave, and truncating to $L$ still preserves concavity. These properties ensure that when the truncated logarithm distance is used to define the loss~\eqref{eq::util_min} and, subsequently, the utility~\eqref{eq::util_num}, submodularity is preserved.

%====================
% Simulations
%====================

\section{Demonstrative examples}
\label{sec::num}
\vspace{-0.05in}
We consider deployment scenarios via a single authority using Algorithm~\ref{alg::SQG_uniform} in the NavEX framework over a $100 \times 100$ workspace $\mathcal{W}$ with targets generated~randomly.

\begin{figure}[t]
    \centering
    \includegraphics[trim=0 8 0 0, clip, width=0.95\linewidth]{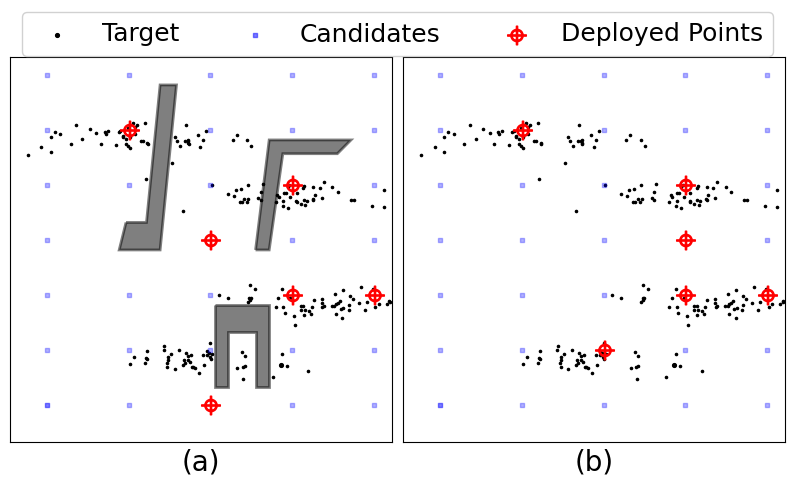}
    \vspace{-0.1in}\caption{{\small Fair-access deployment using NavEX in a workspace with (left plot) and without (right plot) obstacles.}}
    \label{fig:2025_RAL_euclidean_vs_visgraph}
\end{figure}

\begin{figure}[t]
    \centering
    \includegraphics[width=0.7\linewidth]{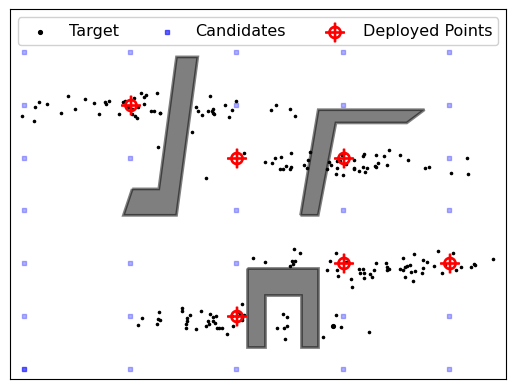}
    \caption{\small{Hotspot deployment using NavEX.} }
    \label{fig:2025_RAL_hotspot_deployment}
\end{figure}

\begin{figure*}[t]
    \centering

    \begin{subfigure}[t]{0.30\linewidth}
        \centering
        \includegraphics[scale=0.128]%width=0.9\textwidth,height=3.50cm
        {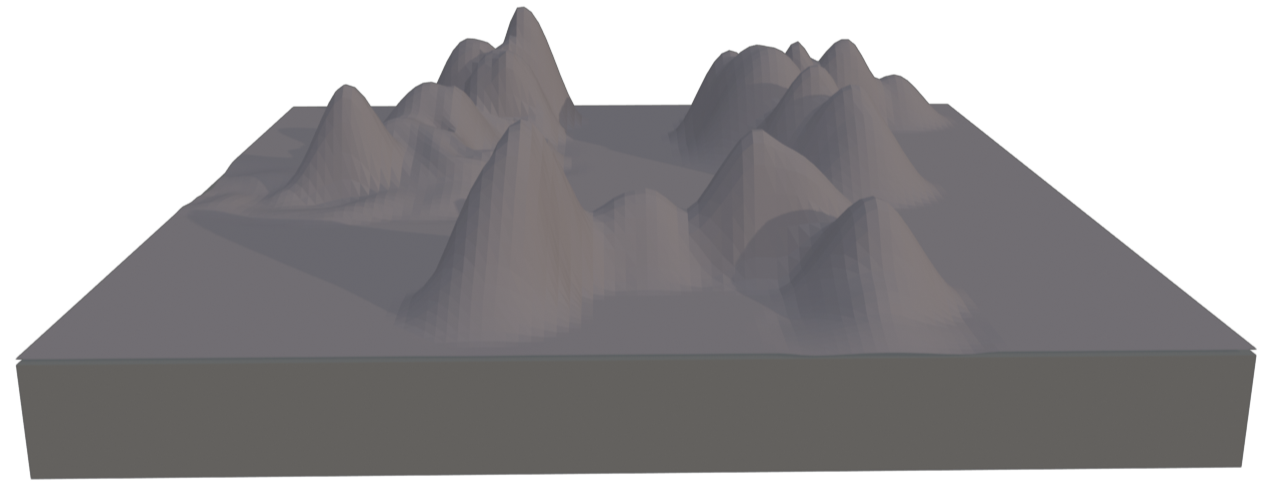}
        \caption{\small{Terrain Model one.}} 
        \label{subfig:2025_RAL_terrain1}
    \end{subfigure}
    \hfill
    % Bottom row of subfigures
    \begin{subfigure}[t]{0.30\linewidth}
        \centering
        \includegraphics[scale=0.22]{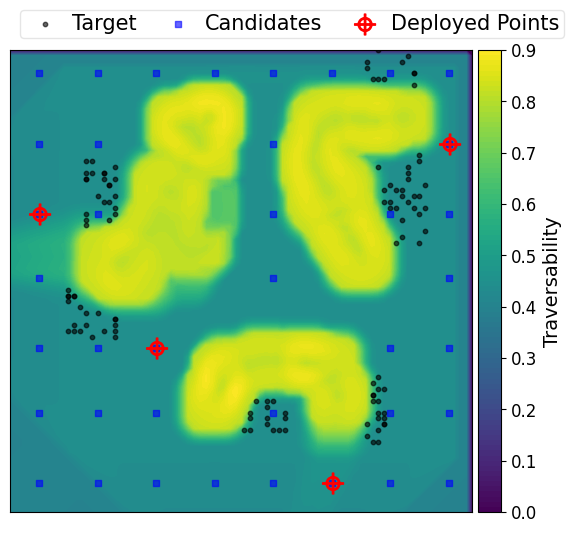}
        \caption{\small Fair-access deployment using traversability-aware RRT$^{\star}$~distance~metric.}
        \label{subfig:2025_RAL_deployment1}
    \end{subfigure}
    \hfill
    \begin{subfigure}[t]{0.30\linewidth}
        \centering
        \includegraphics[scale=0.22]{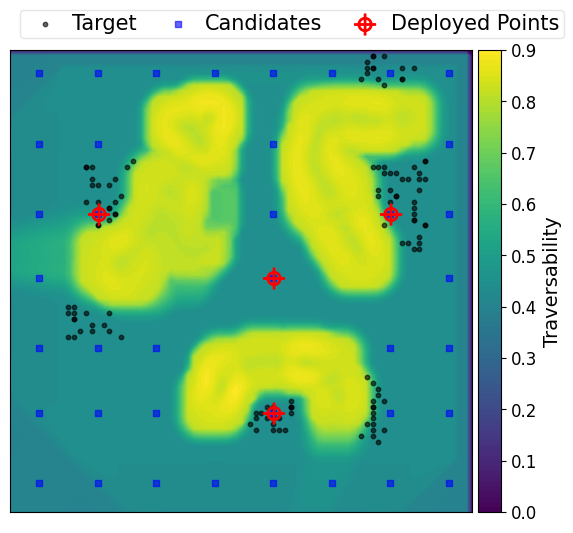}
        \caption{{\small Fair-access deployment using unweighted RRT$^{\star}$ distance metric oblivious to traversability.}}
        \label{subfig:2025_RAL_deployment1_euclidean}
    \end{subfigure}
    
    \vspace{0.5 em}

    % Three subfigures aligned horizontally
    \begin{subfigure}[t]{0.30\linewidth}
        \centering
        \includegraphics[scale=0.1]{Fig_RAL_visibility/2025_visibility_hills_problem_statement.png}
        \caption{Terrain Model two.}
        \label{subfig:2025_RAL_terrain2}
    \end{subfigure}
    \hfill
    \begin{subfigure}[t]{0.30\linewidth}
        \centering
    \includegraphics[scale=0.22]{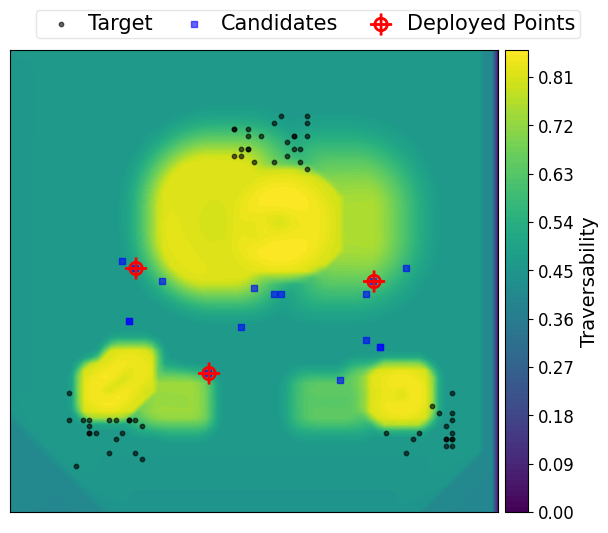}
    \caption{\small Fair-access deployment using traversability-aware~RRT$^{\star}$~distance~metric.}
        \label{subfig:2025_RAL_deployment2}
    \end{subfigure}
    \hfill
    \begin{subfigure}[t]{0.3\linewidth}
        \centering
        \includegraphics[scale=0.22]{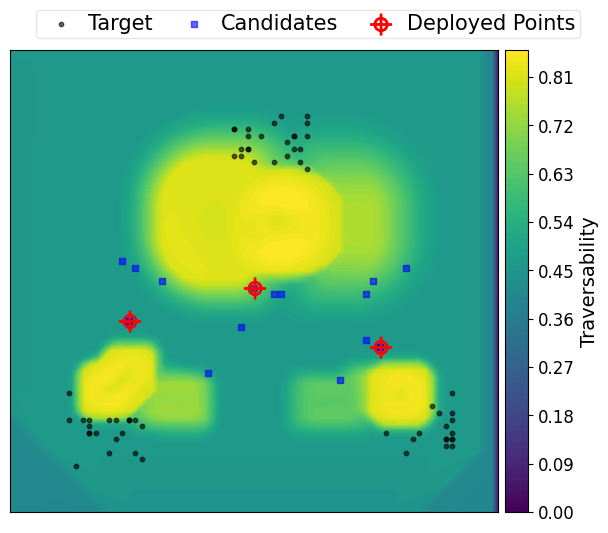}
        \caption{{\small Fair-access deployment using unweighted RRT$^{\star}$ distance metric oblivious to traversability.}}
        \label{subfig:2025_RAL_deployment2_euclidean}
    \end{subfigure}
    \vspace{-0.05in}

    \caption{\small The outcome of fair-access deployment using NavEX in two distinct uneven terrains. Plots (a) and (d) show the terrains, with corresponding deployment scenarios to their right. Heatmaps show traversability maps.}
    \label{fig:2025_RAL_traversability_deployment}
\end{figure*}

\emph{Deployment in 2D Non-Convex Free Workspace}: 
Figure~\ref{fig:2025_RAL_euclidean_vs_visgraph} shows the workspace with and without obstacles and randomly generated targets. In this scenario, six agents are deployed over candidate deployment points, which are grid points in $\mathcal{W}$ as shown in Fig.~\ref{fig:2025_RAL_euclidean_vs_visgraph}. The result from our proposed deployment framework, using the visibility graph-based distance metric, an examplar based clustering and utilizing the sequential greedy Algorithm~\ref{alg::SQG_uniform}, is shown in Fig.~\ref{fig:2025_RAL_euclidean_vs_visgraph}(a). Compared to Fig.~\ref{fig:2025_RAL_euclidean_vs_visgraph}(b), where obstacles are absent and Euclidean distance is used, our algorithm adjusts the deployment points to account for how obstacles affect the shortest distance to the target points.

In fair-access deployment, the objective is equitable access to the deployed agents. We have also simulated the scenario where we aim to deploy our six agents in the same space but with the objective of hotspot deployment. In this case, we use a truncated logarithmic transformation~\ref{eq::hotspot_deploy} and \ref{eq::util_num} to prioritize densely populated areas. Using Algorithm~\ref{alg::SQG_uniform}, agents are deployed to capture the most representative cluster centers, as illustrated in Fig.~\ref{fig:2025_RAL_hotspot_deployment}, effectively aligning with the objective of emphasizing high-priority regions while respecting the spatial constraint imposed by obstacles. The visibility graph is an optimal metric in 2D non-convex domains and preserves the theoretical worst-case guarantee of $(1-\frac{1}{e})OPT$ when using the greedy approach, where OPT is the optimal value.

\emph{Deployment in 3D Simulated Terrain}: 
Next, we evaluated agent deployment for fair-access tasks in two distinct $100 \times 100$ hilly terrains, $\mathcal{W}$, shown in Fig.~\ref{subfig:2025_RAL_terrain1} and Fig.~\ref{subfig:2025_RAL_terrain2}. In these evaluations, we adopt the traversability-aware RRT$^\star$ as our distance metric in NavEX. To maintain feasibility, the candidate deployment points are located within the traversable flat regions. We consider two different terrains. In the terrain depicted in Fig.~\ref{subfig:2025_RAL_terrain1}, the candidate deployment points are grid points scattered evenly within the flat traversable regions. As shown in Fig.~\ref{subfig:2025_RAL_deployment1}, the NavEX framework places the agents in locations that provide equitable access, considering the traversability conditions. In contrast, Fig.~\ref{subfig:2025_RAL_deployment1_euclidean} illustrates that if traversability conditions are ignored, the resulting deployment in practice results in favoring subsets of targets over others or a wasteful deployment (the point in the middle). This outcome occurs because in real-life scenarios, traversability limits the agents' ability to navigate.

In the second scenario, for the terrain depicted in Fig.~\ref{subfig:2025_RAL_terrain2}, the candidate deployment points are located in a region of $\mathcal{W}$ surrounded by hills with varying slopes (traversability). The targets are situated on the opposite side of these hills from the deployment points. Therefore, careful selection of deployment points is crucial to account for slope-dependent traversability. As Fig.~\ref{subfig:2025_RAL_deployment2} illustrates, NavEX places the agents as close as possible to the targets within the traversability-friendly regions. Conversely, when traversability conditions are disregarded, as shown in Fig.~\ref{subfig:2025_RAL_deployment2_euclidean}, the deployment choices are based solely on the shortest geodesic distance (approximated by unweighted RRT$^\star$, oblivious to traversability). In real-world scenarios, this will result in an ineffective deployment due to the physical limitations of the deployed agents in traversing hilly regions.

%====================
% Conclusions
%====================

\section{{Conclusion}}
\vspace{-0.05in}
We have presented NavEX, a novel dispatch coverage framework that provides a robust methodology for multi-agent deployment in non-convex and uneven environments. By combining exemplar-clustering with obstacle-aware and traversability-aware distance metrics, NavEX offers a unified approach to fair-access and hotspot deployment. The submodular optimization foundation of NavEX enables efficient, near-optimal solutions with provable performance guarantees. Simulations demonstrate the framework's effectiveness in complex, realistic settings. Future work will include extending NavEX to heterogeneous agents and integrating real-time adaptability to dynamic conditions like moving targets and changing environments, enhancing the practical relevance for diverse real-world applications.

\vspace{-0.05in}
\bibliographystyle{ieeetr}%
\bibliography{bib/reference.bib, bib/alias.bib}

\begin{thebibliography}{10}

\bibitem{SSK-SM:24}
S.~S. Kia, , and S.~Martinez, ``Multi-agent coverage control: From discrete assignments to continuous multi-agent distribution matching,'' {\em ASME Dynamic Systems \& Control Division Newsletter}, 2024.
\newblock Avaiable at~arXiv:2407.13890.

\bibitem{ME-RG-DD:11}
M.~V. Espina, R.~Grech, D.~De~Jager, P.~Remagnino, L.~Iocchi, L.~Marchetti, D.~Nardi, D.~Monekosso, M.~Nicolescu, and C.~King, {\em Multi-robot Teams for Environmental Monitoring}, pp.~183--209.
\newblock Berlin, Heidelberg: Springer Berlin Heidelberg, 2011.

\bibitem{JC-SM-FB:04}
J.~Cortes, S.~Martinez, T.~Karatas, and F.~Bullo, ``Coverage control for mobile sensing networks,'' {\em IEEE Transactions on Robotics and Automation}, vol.~20, no.~2, pp.~243--255, 2004.

\bibitem{OA-19}
O.~Arslan, ``Statistical coverage control of mobile sensor networks,'' {\em IEEE Transactions on Robotics}, vol.~35, no.~4, pp.~889--908, 2019.

\bibitem{MS-DR-JS:09}
M.~Schwager, D.~Rus, and J.-J. Slotine, ``Decentralized, adaptive coverage control for networked robots,'' {\em The International Journal of Robotics Research}, vol.~28, no.~3, pp.~357--375, 2009.

\bibitem{CN-CH-ZM:08}
C.~H. Caicedo-Nunez and M.~Zefran, ``A coverage algorithm for a class of non-convex regions,'' in {\em {IEEE} Int. Conf. on Decision and Control}, pp.~4244--4249, 2008.

\bibitem{YS-MT-AT:15}
Y.~Stergiopoulos, M.~Thanou, and A.~Tzes, ``Distributed collaborative coverage-control schemes for non-convex domains,'' {\em IEEE Transactions on Automatic Control}, vol.~60, no.~9, pp.~2422--2427, 2015.

\bibitem{SB-RG_VK:14}
S.~Bhattacharya, R.~Ghrist, and V.~Kumar, ``Multi-robot coverage and exploration on riemannian manifolds with boundaries,'' {\em The International Journal of Robotics Research}, vol.~33, no.~1, pp.~113--137, 2014.

\bibitem{JH-FA:20}
J.~Hu, H.~Niu, J.~Carrasco, B.~Lennox, and F.~Arvin, ``Voronoi-based multi-robot autonomous exploration in unknown environments via deep reinforcement learning,'' {\em IEEE Transactions on Vehicular Technology}, vol.~69, no.~12, pp.~14413--14423, 2020.

\bibitem{YA-AA-SS:23}
A.~S. Yengejeh, A.~B. Asghar, and S.~L. Smith, ``Distributed multirobot coverage control of nonconvex environments with guarantees,'' {\em IEEE Transactions on Control of Network Systems}, vol.~10, no.~2, pp.~796--808, 2023.

\bibitem{LK-LK:24}
K.~Lee and K.~Lee, ``Adaptive centroidal voronoi tessellation with agent dropout and reinsertion for multi-agent non-convex area coverage,'' {\em IEEE Access}, vol.~12, pp.~5503--5516, 2024.

\bibitem{LS-OS:16}
S.~Li and O.~Svensson, ``Approximating k-median via pseudo-approximation,'' {\em SIAM Journal on Computing}, vol.~45, no.~2, pp.~530--547, 2016.

\bibitem{RR-LZ-GS:20}
R.~K. Ramachandran, L.~Zhou, J.~A. Preiss, and G.~S. Sukhatme, ``Resilient coverage: Exploring the local-to-global trade-off,'' in {\em IEEE/RSJ Int. Conf. on Intelligent Robots \& Systems}, pp.~11740--11747, 2020.

\bibitem{SX-GC:19}
X.~Sun, C.~G. Cassandras, and X.~Meng, ``Exploiting submodularity to quantify near-optimality in multi-agent coverage problems,'' {\em Automatica}, vol.~100, pp.~349--359, 2019.

\bibitem{AK-AS-CG:08}
A.~Krause, A.~Singh, and C.~Guestrin, ``Near-optimal sensor placements in {G}aussian processes: Theory, efficient algorithms and empirical studies,'' {\em J. Mach. Learn. Res.}, vol.~9, p.~235–284, jun 2008.

\bibitem{kia2025submodular}
S.~S. Kia, ``Submodular maximization subject to uniform and partition matroids: From theory to practical applications and distributed solutions,'' {\em arXiv preprint arXiv:2501.01071}, 2025.

\bibitem{JL-RKW:19}
J.~Liu and R.~K. Williams, ``Submodular optimization for coupled task allocation and intermittent deployment problems,'' {\em IEEE Robotics and Automation Letters}, vol.~4, no.~4, pp.~3169--3176, 2019.

\bibitem{ZX-VT:22}
Z.~Xu and V.~Tzoumas, ``Resource-aware distributed submodular maximization: A paradigm for multi-robot decision-making,'' in {\em {IEEE} Int. Conf. on Decision and Control}, pp.~5959--5966, 2022.

\bibitem{NR-SSK:23auto}
N.~Rezazadeh and S.~S. Kia, ``Distributed strategy selection: A submodular set function maximization approach,'' {\em Automatica}, vol.~153, p.~111000, 2023.

\bibitem{GS-GSS:24}
G.~Shi and G.~S. Sukhatme, ``Inverse submodular maximization with application to human-in-the-loop multi-robot multi-objective coverage control,'' {\em arXiv preprint arXiv:2403.10991}, 2024.

\bibitem{KL-PR:90}
L.~Kaufman and P.~Rousseeuw, {\em Finding Groups in Data: An Introduction To Cluster Analysis}.
\newblock Wiley, 01 1990.

\bibitem{BA-MB-KA:14}
A.~Badanidiyuru, B.~Mirzasoleiman, A.~Karbasi, and A.~Krause, ``Streaming submodular maximization: Massive data summarization on the fly,'' in {\em ACM SIGKDD international conference on Knowledge discovery and data mining}, pp.~671--680, 2014.

\bibitem{MD:00}
M.~De~Berg, {\em Computational geometry: algorithms and applications}.
\newblock Springer Science \& Business Media, 2000.

\bibitem{JG-SS-TB:14}
J.~D. Gammell, S.~S. Srinivasa, and T.~D. Barfoot, ``Informed rrt*: Optimal sampling-based path planning focused via direct sampling of an admissible ellipsoidal heuristic,'' in {\em IEEE/RSJ Int. Conf. on Intelligent Robots \& Systems}, pp.~2997--3004, 2014.

\bibitem{LP-MW:79}
T.~Lozano-P\'{e}rez and M.~A. Wesley, ``An algorithm for planning collision-free paths among polyhedral obstacles,'' {\em Commun. ACM}, vol.~22, p.~560–570, Oct. 1979.

\bibitem{SG:07}
S.~K. Ghosh, {\em Visibility Graphs}, p.~136–170.
\newblock Cambridge University Press, 2007.

\bibitem{SK-EF:11}
S.~Karaman and E.~Frazzoli, ``Sampling-based algorithms for optimal motion planning,'' {\em The International Journal of Robotics Research}, vol.~30, pp.~846 -- 894, 2011.

\bibitem{AL-ML-LL:24}
A.~Leininger, M.~Ali, H.~Jardali, and L.~Liu, ``Gaussian process-based traversability analysis for terrain mapless navigation,'' in {\em {IEEE} Int. Conf. on Robotics and Automation}, pp.~10925--10931, 2024.

\bibitem{RG-AK:10}
R.~Gomes and A.~Krause, ``Budgeted nonparametric learning from data streams,'' in {\em International Conference on Machine Learning}, (Haifa, Israel), 2010.

\end{thebibliography}

\end{document}